\documentclass[conference]{IEEEtran}
\IEEEoverridecommandlockouts
\usepackage{cite}
\usepackage{caption} 
\usepackage{amsmath,amssymb,amsfonts}
\usepackage{pdfpages}
\usepackage{booktabs}
\usepackage{textcomp}
\usepackage{xcolor}
\usepackage{tablefootnote}
\usepackage[misc]{ifsym}
\usepackage{caption}
\usepackage{float} 
\usepackage{subcaption}
\usepackage{adjustbox}
\usepackage{subcaption}
\usepackage{amsmath,amssymb,amsfonts}
\usepackage{amsthm}
\usepackage{multirow}
\usepackage{bibentry}
\usepackage{listings} 
\usepackage{float}
\usepackage{color}
\usepackage{url}
\usepackage{enumerate}
\usepackage[ruled,linesnumbered]{algorithm2e}
\usepackage{bm}
\usepackage{stfloats}
\usepackage{graphicx}
\usepackage{epsfig}
\usepackage{epstopdf} 
\usepackage[T1]{fontenc}
\usepackage[colorlinks,
linkcolor=black,
anchorcolor=black,
citecolor=black
]{hyperref}

\ifCLASSOPTIONcompsoc

  \usepackage[nocompress]{cite}
\else

  \usepackage{cite}
\fi

\ifCLASSINFOpdf

\else

\fi

\newtheoremstyle{custom}
  {\topsep}   
  {\topsep}   
  {\itshape}  
  {}          
  {\bfseries} 
  {:}         
  {.5em}      
  {}          

\theoremstyle{custom}


\renewcommand{\proofname}{Proof}

\begin{document}

\title{Two-Timescale Digital Twin Assisted Model Interference and Retraining over Wireless Network
}

\author{\IEEEauthorblockN{
Jiayi Cong$^{\star,\dagger}$, 
     Guoliang Cheng$^{\star,\dagger}$, Changsheng You$^{\dagger}$, Xinyu Huang$^{\ddagger}$, and Wen Wu$^{\star}$}
      
        \IEEEauthorblockA{
    Frontier Research Center, Peng Cheng Laboratory, Shenzhen, China$^\star$\\
  The Department of Electronic and Electrical Engineering, Southern University of Science and Technology$^\dagger$\\
  The Department of Electrical and Computer Engineering, University of Waterloo$^\ddagger$\\
     Email:  \mbox{\{congjy, chenggl, wuw02\}@pcl.ac.cn}, \mbox{youcs@sustech.edu.cn}, and \mbox{x357huan@uwaterloo.ca}
    }
}

\maketitle

\begin{abstract}
In this paper, we investigate a resource allocation and model retraining problem for dynamic wireless networks by utilizing incremental learning, in which the digital twin (DT) scheme is employed for decision making. A two-timescale framework is proposed for computation resource allocation, mobile user association, and incremental training of user models. To obtain an optimal resource allocation and incremental learning policy, we propose an efficient two-timescale scheme based on hybrid DT-physical architecture with the objective to minimize long-term system delay. Specifically, in the large-timescale, base stations will update the user association and implement incremental learning decisions based on statistical state information from the DT system. Then, in the short timescale, an effective computation resource allocation and incremental learning data generated from the DT system is designed based on deep reinforcement learning (DRL), thus reducing the network system's delay in data transmission, data computation, and model retraining steps. Simulation results demonstrate the effectiveness of the proposed two-timescale scheme compared with benchmark schemes.
\end{abstract}

\section{Introduction}

Network virtualization and native artificial intelligence (AI) technology 
have been recognized as a reasonable technology to provide a wireless network that adapts to diversified quality-of-service (QoS) for mobile users (MUs) \cite{ShenSurvey}. In particular, massive user access and extremely low latency requirements in network pose unprecedented opportunities and challenges. For example, autonomous driving has a stringent delay requirement with extreme mobility, e.g., less than 100\textit{ms} \cite{newiov}. In addition, with the development of AI technology, users will have brand new services, such as image identification and natural language generation, which require to be processed.

Specifically, when operating with MUs, the mathematical optimization method cannot be realized due to the unbearable computation cost \cite{newmath}. In addition, real-time resource allocation in the physical world can be criticized for security and reliability, and complex dynamic network scenarios leads to significant system delay performance degradation. Therefore, it is necessary to explore innovative methods to develop resource allocation strategies for MUs. As a remedy to these limitations, a promising solution is to explore AI and network virtualization into resource allocation, in which the MU can offloading these computation tasks to base stations (BSs) for prompt processing, which needs to allocate the computation resources carefully and adapt to the dynamic scenario variation \cite{TwoTimeDT}. 

Digital twin (DT), having real-time interaction with physical entities and creating high-fidelity virtual world, is expected to become a key technology for the transformation of the physical and digital world. One reasonable paradigm of DT  is generating high-quality data in a digital environment based on AI models, e.g. deep neural networks (DNN), to assist the training/retraining of AI models \cite{DataDT}. In addition, DT can also achieve efficient task offloading in dynamic environments and reduce offload latency \cite{newcypertwin}. However, the introduction of DT technology may still lead to performance decline, due to massive data processing and the real-time state updates, especially in the model retraining phase. One reasonable solution is to execute an online model update scheme. By utilizing this scheme, the amount of data required for the DT update and migration can be reduced \cite{InformationDataDT}. While the existing works focus on model retraining based on truthful data, the statistical information of data is yet to be considered in the DT system and real time model update will lead to unnecessary computation cost.

Promising retraining technologies, such as continual learning and continue learning, are proposed for real-time processing and efficient model retraining \cite{IMSURVEY}. However, fast model switching is difficult to achieve considering delay-tolerance constraints. Some features in wireless communication scenarios have statistical properties, e.g., the MU traffic and captured images distribution, show a certain regularity on a large-timescale \cite{TongLiTraffic}, which brings new opportunities. When MUs travel to another scenario, BSs are able to support MUs in retraining the image identification model which can adapt to the new distribution of captured images. In order to reduce the system delay, MUs may not be retrained in every slot, while the computation resources needs to be allocated in every slot suitably. Hence, a comprehensive network management policy should jointly consider resource allocation and retraining decision in different timescales.

In this paper we propose a two-timescale DT-based resource allocation scheme. An optimization problem to minimize long-term system delay is formulated, by optimizing user association, computation resource allocation, and incremental learning decisions. Specifically, in the large timescale, DT updates and makes user association and incremental learning decisions by jointly considering statistical information in the DT system and real-time states in the physical network. In the short timescale, BSs resource allocation and data-allocation decisions are leveraged under the resource and model accuracy constraints. The main contributions of this paper are summarized as follows:
\begin{itemize}
    \item We present a two-timescale framework for dynamic networks to support each MU model interference and retraining, and balance BSs’ computation resources;
    \item We formulate the two-timescale framework as a long-term expectation optimization based problem. The objective is to dynamically make resource allocation, user association, and incremental learning decisions to minimize system delay while satisfying coupled constraints;
    \item We design a DT-based decision making and data generation system within the scheme to achieve higher model accuracy without significantly increasing system delay, while satisfying resource constraints at each BS.
\end{itemize}

The remainder of this paper is organized as follows. We present the system model of physical and DT environment in Section II. Section III provides the large-timescale and short-timescale algorithms in detail. Simulation results are then presented in Section IV. Finally, we give the conclusion and future directions in Section V.

\section{System Model and Problem Formulation}

\subsection{Considered Scenario}

\begin{figure}[t]
		\centering
		\includegraphics[width=9cm]{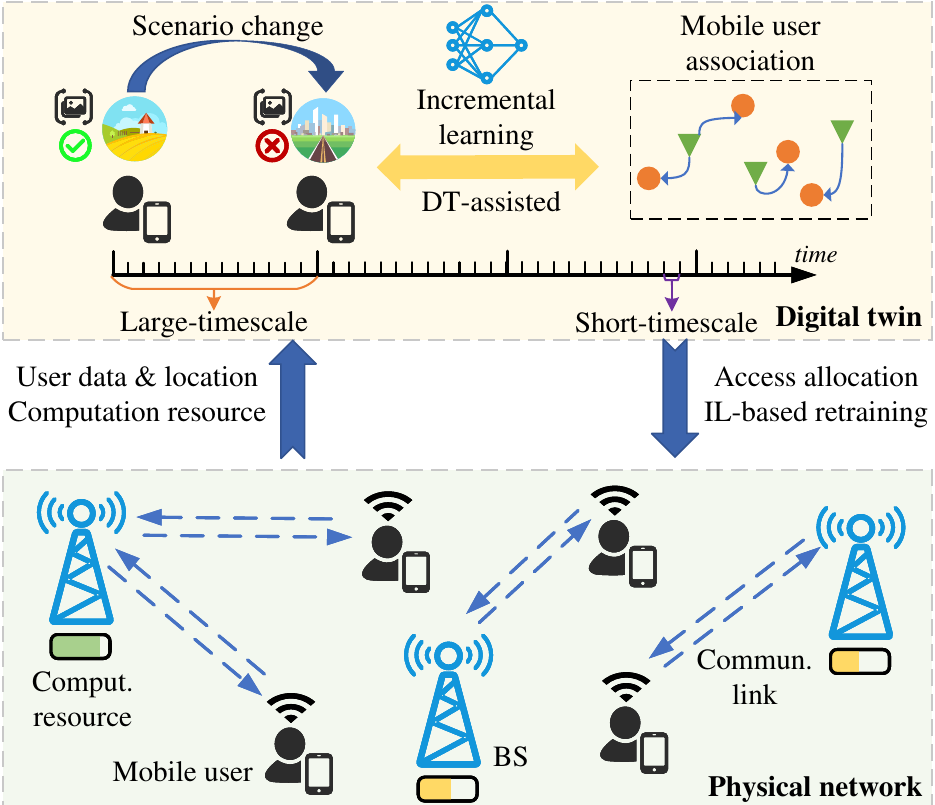}
		\caption{Architecture of DT-assisted networking system.}
		\label{SM}
	\end{figure}

In this section, we design a DT-assisted mobile network architecture for the efficient computation and communication resource allocation. This architecture considers the image recognition tasks for MUs, as shown in Fig.~\ref{SM}. Specifically, the network architecture consists of $M$ BSs in set $\mathcal{M}=\{1,2,\dots,M\}$, and $N$ MUs in set  $\mathcal{N}=\{1,2,\dots,N\}$, operating in a time-slotted manner.

In the scenario, AI models are deployed in the DT to make communication and computation resource allocation decisions for the physical networks. The BS side, with abundant computation resources compared to the user side, can assist MUs in computation tasks by processing the upload task from MUs with insufficient computation resources and allocating the available frequency band.
Each MU possesses a local requirement such as image classification. 
In this DT assisted mobile network system, the distribution of MUs changes over time and different MUs have their personalized computation requirements, in which the AI model, primarily responsible for making communication and computation resource allocation decisions, operates in the digital twin layer and sends them to the physical world.

We focus on BS side resource allocation problem. Considering a discrete-time system in which the total time length $T_s$ is divided into several unit \textit{time slot} denoted as $\mathcal{T} \triangleq \{1,...,T\}$ and $T_s = Tt_s$. Since the MU association maintain a steady state for a certain period of time, we further equally divide the considering time into \textit{time frame} denoted as $T^f = St_s$ where $S$ is the number of unit time in each time frame and $S>1$.

\subsection{Network System}

 Let $\mathcal{D}^\text{BS}= \{(x_m, y_m)\}$ denote the location of each BS, where $x$ and $y$ denote the x- and y-coordinate in the physical world Cartesian coordinate system, respectively. Consider that the movement of MUs typically exhibits temporal dependency, namely, their current position and computation requirement as  $\mathcal{D}_t= \{(x_{n}^t, y_{n}^t)\}$. Then the distance between the $m$-th BS and $n$-th in time slot $t$ is shown as $d_{m,n}^t = ||d^{\text{BS}}_m-d_n^t||_2$. Let $v_{m,n}^t$ represent the BS-MU connection indicator. Especially, $v_{m,n}^t = 1$ if
the $m$-th BS connected with $n$-th MU through wireless link; otherwise, $v_{m,n}^t = 0$. In order to ensure that each user can create and maintain their own digital twin, MU needs to connect with least one BS at each time slot, even if its computation task do not require BS assistance, i.e., $\sum_{m=1}^M {v_{m,n}^t} \geq 1, n\in \mathcal{N}_t, t\in \mathcal{T}$. Let the $h_{m,n}^t$ denote the channel gain of each pair of BS and MU in time slot $t$. For each MU's individual computation task, defined as $Q^t_n$, can be divided into two parts: locally computation tasks $Q^\text{L}_n$ and BS computation tasks $Q^t_{m,n}$ where $Q^\text{L}_n + \sum_{m=1}^M Q^t_{m,n} = Q^t_n$. In each time slot, MUs will update the user location and system delay of personalized tasks completed through the established wireless connection between BS-MU pairs, which includes both large-scale and small-scale fading. Since orthogonal frequency division multiplexing (OFDM) technology is employed in this system, the communication interference between different MUs is neglected, and Gaussian noise is considered as interference in the physical wireless network.

\subsubsection{Communications Delay} For the uplink transmission, the signal-to-noise ratio (SNR) received at BS $m$ is given by 
\begin{equation}
   r_{m,n}^t = \frac{\omega_{m,n}^th_{m,n}^t}{\sigma^2},
\end{equation}
where $\omega_{m,n}^t$ is the transmit power of each MU in time slot $t$, and $\sigma^2$ is the noise power. 

Then the transmit rate of the uplink transmission rate $R_{m,n}^t$ can be calculated as
\begin{equation}
   R_{m,n}^t = Bv_{m,n}^t\log_2(1+ r_{m,n}^t),
\end{equation}
where $B$ is the spectrum bandwidth. Each allocated task's transmission time-delay is shown below
\begin{equation}
   L_{m,n}^{\text{Trans},t}= 
   \begin{cases}
      \frac{Q^t_{m,n}+Q^\text{D}}{R_{m,n}^t}, &v_{m,n}^t = 1,\\
       ~~~~~0~~~~, &v_{m,n}^t = 0,
   \end{cases}
\end{equation}
where $Q^\text{D}$ is the downlink transmission data. Since the computation task usually needs results or model parameters, e.g., data analyses and AI model training, the downlink delay can be considered as fixation data size. Specially, even if the task is too large to be uploaded in one time slot, it can be divided into different time slots for uploading, so the above constraint is reasonable and realizable. 


\subsubsection{Computation Delay} Considering the MUs lack sufficient computation resources to train/update the model, MUs will upload the obtained data to the BS for storage and training. Although the MU does not need to update the model in the current time slot, in the DT-assisted system we considered, the DT can store data and help improve the accuracy of the future model retraining.

For the locally task computation, the system delay of computation can be formulated as
\begin{equation}
   L_{n}^{\text{L},t} = {Q^\text{L}_n}{f}^{\text{L}},
\end{equation}
where ${f}^\text{L}$ is the locally computation ability which is assumed as a constant for each MU. For the uploading tasks, the computation delay of each uploading task from MUs sets $\mathcal{N}$ in BS $m$ can be shown as
\begin{equation}
   L_{m,n}^{\text{BS},t} = \frac{Q^t_{m,n}}{{\lambda}^{\text{BS},t}_{m,n}},
\end{equation}
where ${\lambda}^{\text{BS},t}_{m,n}$ is the allocated BS computation resource, which is related to the working frequency and the number of floating point
operations processed in one cycle of the BS proposed in \cite{ZuguangConf}. Considering that the total computation resource of each BS
is limited in each short-timescale, then the computation resource constrain can be formulated as:
\begin{equation}
{\lambda}^{\text{BS},t}_{m} + \sum_{n=1}^N {\lambda}^{\text{BS},t}_{m,n} \leq {\lambda}^{\text{BS}}_{m}.
\end{equation}

As the BS side adopts the parallel processing pattern, the total delay of uploading task for the $m$-th BS can be calculation as
\begin{equation}
   L_{m}^{t} =  \mathop{\max}\limits_{n \in \mathcal{N}}\{L_{m,n}^{\text{BS},t} + L_{m,n}^{\text{Trans},t}\},
\end{equation}
which includes communication and computation delay.

\subsubsection{Digital Twin Assisted Model Retraining Delay}

To enhance the performance of the performance of the dynamic networking in physical world, we propose a scheme architecture incorporates DTs into dynamic networking. DTs can predict the future state of network by modeling the statistical characteristics of historical state/data to ensure that the model meets the accuracy requirements on a large-timescale. The accuracy of each MU is $c_{n}^{t}$, and the threshold of accuracy is $\bar{C}$. 

After obtain the accuracy of the existed model, the DT-based incremental learning scheme is proposed to improve the model accuracy and system robustness by update the existed model. Define binary variable $I^{t^f} \in \{0,1\}$, $I^{t^f} = 1$ if
the $C_{n}^{t^f} \leq \bar{C}$ and indicating the model needs be updated; otherwise, $C_{n}^{t^f}= 0$. Although incremental learning can use small samples to update existing models, more high-quality data can effectively improve the performance of new models after retraining, which can increase the model duration time and reduce system delay. The DT system generated data which supplied the model retraining will related the real-time request data sets $Q^t_{m,n}$. In dynamic scenarios, the Kullback-Leibler (KL) divergence relationship between the actual data sets and the DT-stored data sets are considered to determine the optimal DT-generated data distribution probability, which can be shown as
\begin{equation}
   p^t_{m,n}(x) =  \mathop{\arg\min}\limits_{p\in\mathcal{P}} \sum_{x\in Q^t_{m,n}}p^t_{m,n}(x)\log\frac{p^t_{m,n}(x)}{q^t_{m,n}(x)},
   \label{DTdata}
\end{equation}
where $\mathcal{P}$ is discrete executable set of DT-generated data distribution probability $p$ and $q^t_{m,n}(x)$ is the distribution of real data under DT classification. To guarantee that the newly DT-generated data can be quickly adjusted in current scenario, the distribution probability with the current data distribution, i.e., $p_{m,n} \leftarrow p^t_{m,n}$. Then, the DT generated data size $S^t_{m,n}$ is formulated as follow
\begin{equation}
   S^t_{m,n} =  \delta_{m,n}^t p^t_{m,n}||Q^t_{m,n}||_2,
   \label{DTdataset}
\end{equation}
where $\delta_{m,n}^t$ is the  size of the data generated by DT relative to the data obtained in real time resulting in additional computation delay. This part of computation delay can be calculated similarly with the computation model of physical system as
\begin{equation}
   L_{m,n}^{\text{DT},t} = \frac{S^t_{m,n}}{{\lambda}^{\text{BS},t}_{m,n}}.
\end{equation}

Then the the effective delay is $L_{n}^{t}+L_{m,n}^{\text{DT},t}$ where the $L_{m,n}^{\text{DT},t}$ may equal zero and the minimize delay problem will degrade into a communication-computation trade off problem. The incremental learning is utilized to update existing models by using efficient solution, e.g., the mini-batch stochastic gradient descent \cite{HuangIL}.

\subsection{Problem Formulation}
Due to temporal variations in the model data distribution, which result in decreased model accuracy, it is critical to minimize the overall system delay over the long-term while meeting resource and association constraints, particularly from a network operator's perspective. The optimization problem, denoted as \textbf{P1}, can be mathematically formulated as:
\begin{subequations}
 	\begin{align}
 		({\bf P1}):\min_{v_{m,n}^t, c_{m,n}^t }  
            &\mathbb{E}\{\min_{\substack{Q^t_{m,n}, {\lambda}^{\text{BS},t}_{m,n}} }\frac{1}{N_t}\sum_{n=1}^{N_t} L_{n}^{t}+c_{m,n}^tL_{m,n}^{\text{DT},t}\} \label{P1}\\
 		\text{s.t.}~
            &~ c_{m,n}^t \in \{0, 1\}, ~\forall m,n \label{P1b}\\
 		&~ v_{m,n}^t \in \{0, 1\}, ~\forall m,n  \label{P1c}\\
            &~ 1 \leq \sum_{m=1}^M {v_{m,n}^t} \leq M, ~\forall n \label{P1d}\\
 		&~Q^\text{L}_n + \sum_{m=1}^M Q^t_{m,n} = Q^t_n, ~\forall n\label{P1e}\\
            &~	{\lambda}^{\text{BS},t}_{m} + \sum_{n=1}^{N_t} {\lambda}^{\text{BS},t}_{m,n} \leq {\lambda}^{\text{BS}}_{m}, \label{P1f}
 	\end{align}
 \end{subequations}
 where \eqref{P1b} is the incremental learning indicator constraint and represents the decision for model retraining. \eqref{P1c} represents the binary constraint on the association relationship variables. \eqref{P1d} restricts that, during each time frame, at least one BS can be assigned to associate to each MU. \eqref{P1e} indicates that each device needs to complete all data transmission task in each time slot. \eqref{P1f} represents the computation resource constraint for each BS in a single time slot. In summary, within the resource and allocation constraints, the objective of the DT-assisted model incremental retraining problem is to minimize the average delay. 
 
 However, the problem is a mixed integer nonlinear programming problem which is a nondeterministic polynomial (NP)-hard problem and needs decisions across different timescales. To address this issue, we propose a two-timescale method which can ensure consistency and comparability in different timescales.

\section{Proposed Algorithm}
Given the dynamic nature of the network environment and the extensive solution space of the optimization problem, which requires real-time decision making, we employ deep reinforcement learning (DRL) to address the allocation problem in each time slot. Furthermore, for the large-timescale, a model retraining scheme and user association based on DT and statistical information are proposed to achieve high robustness.

\subsection{Large-Timescale Algorithm}
At each time frame $t\in [1,st_s]$, the BS first acquires the effective channel and location of each MU. With the fixed historical data of MU model accuracy situation and transmission time, the BS designs the large-timescale access allocation for each user $v_{m,n}^t$. Furthermore, we relax the amplitudes of $v_{m,n}^t$ to be in the interval $[0,1]$, which is shown to help accelerate the convergence of the proposed algorithm by simulation in this section as shown in the following problem

\begin{subequations}
 	\begin{align}
 		({\bf P2}):\min_{v_{m,n}^t}  
            &~ \eta \label{P2}\\
 		\text{s.t.}~
 		&~ \mathbb{E}\{\frac{1}{N_t}\sum_{n=1}^{N_t} L_{n}^{t}+c_{m,n}^tL_{m,n}^{\text{DT},t}\} \leq \eta\\
            &~ \eqref{P1b}~\text{and}~\eqref{P1c}, \label{P2b}
 	\end{align}
 \end{subequations}
where $\eta$ is the introduced slack variable related with \eqref{P1} with fixed incremental learning indicator $c_{m,n}^t$. \textbf{(P2)} is standard linear programming can be solved using existing liner programming tools or the simplex method. To get the optimal integer solution for \textbf{(P2)}, we denote the optimal integer solution of the relaxed problem is $\hat{v}_{m,n}^t = \{\hat{v}_{m,n}^t, \forall m,n\}$ and the original optimal problem solution is ${v}_{m,n}^t = \{v_{m,n}^t, \forall m,n\}$ and Branch-and Bound method with the obtained optimal solution is proposed and the detailed process is presented in Algorithm~\ref{alg1} which can be solved by existed toolbox, i.e., CVX \cite{CVX}.
	
After obtaining the access allocation $v_{m,n}^t$, the DT-assisted incremental learning optimization decision problem is to ensure that every MU satisfies the accuracy constraint in the time frame, which can be shown as \eqref{accuracy_2}. Since different MUs make incremental learning decisions independently, the decision-making process can be solved independently.

\begin{algorithm}[t]\small
\caption{Proposed user association algorithm}
	\label{alg1} 
\KwData{Obtain initialize $\hat{v}_{m,n}^t$ by solving \eqref{P2} with random ${Q^t_{m,n}, {\lambda}^{\text{BS},t}_{m,n}}$  and initialize $m$=1.}
\KwResult{Optimal MU association ${v}_{m,n}^t$}
\While{$m \leq M$}{
solve \eqref{P2} with Branch and Bound method\;
\eIf{$\hat{v}_{m,n}^t$ are integer solutions}{
      Set ${v}_{m,n}^t = \hat{v}_{m,n}^t$\;
      }{
      Find potential solutions in $\{\hat{v}_{m,n}^t\}$ and solve \eqref{P2}, respectively\;
      Find solutions with maximum optimal value and set {${v}_{m,n}^t = \hat{v}_{m,n}^t$}\;
      }
$m = m+1$;}
\end{algorithm}

Since a sudden decrease in accuracy over a time slot due to a sudden boost cannot be relied upon, accuracy over a time frame can be expressed as
\begin{equation}
   c_{n}^{t^f} =  \frac{\sum_{St^f}^{S(t^f+1)}{c_{n}^{t}}}{T^f}.
   \label{accuracy}
\end{equation}

While \eqref{accuracy} is a posterior formula, the calculated long-term accuracy will be obtained at $t^f+1$. In DT system, we can obtain the statistical probability model accuracy related with user location and model using time of duration which can be shown as $f_c(t|\mathcal{D}_t,{T}^{\text{dur}}_n)$ where ${T}^{\text{dur}}_n$ is the model using time of duration for different MUs. Then \eqref{accuracy} can be calculated as
\begin{equation}
   C_{n}^{t^f} =  \frac{\int_{t_s}^{T^f} f_c(t)~\text{d}t}{T^f}.
   \label{accuracy_2}
\end{equation}

Moreover, each considered MU has its own retraining decision variable, which only related with the considering time window and accuracy threshold $\bar{C}$. The model retraining indicator yields the following binary decision:

\begin{equation}
   c_{m,n}^t= 
   \begin{cases}
      1, &C_{n}^{t^f} \leq  \bar{C}     \\
      0, &C_{n}^{t^f} \geq \bar{C}
   \end{cases}.
\end{equation}

By utilizing DT technology, the decision-making is implemented using historical data, and the incremental learning data is generated based on the statistical distribution in the DT system.

\subsection{Short-Timescale Algorithm}
After obtaining the large-timescale decision which affects for a period of short-timescale, for time slot $t\in[1,t_s]$, the BSs will allocate the computation resources and MUs will select the data and transmit data to the BS for training and storage. The short-timescale problem can be rewritten as
\begin{subequations}
 	\begin{align}
 		({\bf P3}): 
            &~\min_{\substack{Q^t_{m,n}, {\lambda}^{\text{BS},t}_{m,n}} } L_{n}^{t}+c_{m,n}^tL_{m,n}^{\text{DT},t} \label{P3}\\
 		\text{s.t.}~
            &~ \eqref{P1e}~\text{and}~\eqref{P1f}, \label{P3a}
 	\end{align}
 \end{subequations}
while the long-term expectations translates into the minimization of each short-timescale system delay.

The above problem (\textbf{P3}) is difficult to solve directly due to the extremely large continuous variable space for both $Q^t_{m,n}$
 and ${\lambda}^{\text{BS},t}_{m,n}$ and coupled optimization variables. In this paper, we utilize the DRL \cite{DTDRL} method for solving the \eqref{P3}. Firstly, the parameters are described as follows based on the defined Markov decision process (MDP). A DRL can be denoted as $\mathcal{<S, A, R, \pi>}$, which is composed of state space $S$, action space $A$, reward function $R$, and policy $\pi$. The time steps are used alternated and elements are formulated as
\begin{algorithm}[t]\small
\caption{Two-timescale scheme for the DT-assisted network model retraining algorithm}
	\label{alg2} 
\KwData{Large-timescale count number $S$, the information of BS $M, \mathcal{D}^\text{BS}, {\lambda}^\text{BS}$ and the information of MU $N, {h}_{m,n}, {Q}_{n}$.}
\KwResult{computation resource allocation and retraining model.}

\For{each small-time slot $t$}{
\eIf{$\text{mod}(t,S)==0$}{
$\rhd$ Large-timescale problem\;
Solve the large-timescale problem by \textbf{Algorithm \ref{alg1}}\;
}{
$\rhd$ Short-timescale problem\;
\For{$m\in \mathcal{M}$ and $n\in \mathcal{N}$}{
\If{$c_{m,n}^t == 1$}{
$\rhd$ DT-assisted data generation\;
Generate the DT-assisted incremental learning data through \eqref{DTdata}, \eqref{DTdataset}\;}
\For{each episode $\tau = 1,...,\tau_\text{episode}$}{
Choose and execute action $\mathbf{A}(\tau)$, calculate $\mathbf{R}(\tau)$ with (15) and receive $\mathbf{S}(\tau+1)$\;
Store $<\mathbf{S}(\tau), \mathbf{A}(\tau), \mathbf{R}(\tau), \mathbf{S}(\tau+1)>$ in $\mathcal{B}$\;
Update the DRL network parameters $w^{\pi}$\;
}
}
}
Clear experience buffer $\mathcal{B}$\;
}
\end{algorithm}

\begin{enumerate}[1)]
\item \textbf{\textit{State:~}}By observing the dynamic network environment that DT constructed by the DT, the state of the network at time slot $t$ consist of available BSs computation resource $\bm{{\lambda}}^{\text{BS}}(t) = \{{{\lambda}}^{\text{BS}}_{1,1},{{\lambda}}^{\text{BS}}_{1,2},...,{{\lambda}}^{\text{BS}}_{m,n}\}$, MUs to be processed data $\mathbf{{Q}}(t) = \{{Q}_{1}(t) ,{Q}_{2}(t) ,...,{Q}_{n}(t) \}$ and DT-generated data distribution $\mathbf{{p}}(t) = \{{p}_{1,1}(t) ,{p}_{1,2}(t) ,...,{p}_{m,n}(t)\}$ can be expressed as
\begin{equation}
   \mathbf{S}(t) =  \{\bm{{\lambda}}^{\text{BS}}(t),\mathbf{Q}(t),\mathbf{p}(t)\}.
\end{equation}
\item \textbf{\textit{Action:~}}After receiving the current state information, the BS needs to allocate the computation resource and MUs allocate the set of transmission data. The actions of the system can be denoted by
\begin{equation}
   \mathbf{A}(t) =  \{Q^t_{m,n}, {\lambda}^{\text{BS},t}_{m,n}, \forall m,n \}.
\end{equation}
\item \textbf{\textit{Reward:~}}The system utilizes the reward function $r$ to evaluate the action. In each round, the agent responsible for the device selection adopts action $\mathbf{A}$ in state $\mathbf{S}$. The reward function is formulated as follows:
\begin{equation}
   \mathbf{R}(t) = \delta_\text{Q}\sum_{n=1}^N\sum_{m=1}^M (Q^t_n - Q^t_{m,n})- L_{n}^{t}-\alpha_{m,n}^tL_{m,n}^{\text{DT},t},
   \label{reward}
\end{equation}
where the first summation class is the penalty term indicated that the constraint of \eqref{P3a} and the computation resource constraint is considered in the second part of \eqref{reward}.

\item \textbf{\textit{Policy:~}}The policy $\mathcal{\{\pi:S\leftarrow A\}}$ denotes the mapping between the state space and the action space. In each round, the executed action is determined through strategy $\mathbf{A}(t) = \pi(\mathbf{S}(t))$. The DRL explores policy and value functions through DNN, which are considered as the most effective method to solve complex MDP models. We utilize this type of DRL policy in \cite{MushuDQN}.
\end{enumerate}

This scheme exploits the replay buffer mechanism in each training step to ensure that the training data is independently and identically distributed. The complete algorithm for the proposed two-timescale resource allocation as Algorithm~\ref{alg2}.


\section{Simulation Results}

\subsection{Simulation Setting}
In this section, we evaluate the performance of the proposed DT-based incremental learning for an image identification task. For the experimental evaluations, we focus on the image classification task using the CIFAR-10 dataset \cite{krizhevsky2009learning}. 
The channel information is simulated using the Phased Array System Toolbox of MATLAB and the basic DNN model ResNet-18 is selected, and the computation cost of using it for processing each task is considered by \cite{epoch2022estimatingtrainingcompute}. The details of the experiment settings is shown as Table~\ref{tab1}. All numerical experiments are conducted using Python 3.9 with PyTorch on an Intel Core i9-13900HX workstation with 32GB of RAM and GPU involvement.

\begin{table}[t]\small
\renewcommand\arraystretch{1.1}
\centering
\caption{Simulation Parameters.}
\label{tab1}
\setlength{\tabcolsep}{7mm}{
\begin{tabular}{c|c}
\toprule
\textbf{Parameter}                               & \textbf{Value}         \\ \hline\hline
Training samples                                 & 40,000        \\
Test samples                                     & 10,000        \\
Incremental samples                              & 5,000       \\
The number of BS, $M$                             & 3          \\
The number of MU, $N$                             & $U(5, 10)$        \\
Unit time, $t_s$                                  & 5ms         \\
Time slot count, $T$                              & 2          \\
Time frame count, $S$                             & 10         \\
Accuracy constraint, $\bar{C}$                    & 0.85          \\
Computation resource, ${\lambda}^{\text{BS}}$ \cite{epoch2022estimatingtrainingcompute}   & 50 GFLOPs       \\
MU transmission data size, $Q^t_{m,n}$            & $U(10, 20)$     MB       \\
Transmission power, $\omega$                           & -10 dBm           \\ 
Transmission noise power, $\sigma^2$              & -100 dBm            \\ 
\bottomrule
\end{tabular}
}
\end{table}

We adopt the following three benchmark schemes:
\begin{itemize}
    \item \textbf{Without incremental learning}: In this scheme, the system updates the DNN model by retraining it totally with new data collected from the physical environment only;
    \item \textbf{Without DT}: In this scheme, the system's incremental learning data is exclusively derived from the physical environment, and resource allocation is designed for each large timescale;
    \item \textbf{Single timescale case}: This scheme implements all decisions within each small time slot. It is important to note that if the constraint is not met, the system's delay cannot be guaranteed and is considered as break out.
\end{itemize}

\subsection{Simulation Results}

\begin{figure*}[htbp]
	\centering
	\begin{subfigure}{0.325\linewidth}
		\centering
		\includegraphics[width=1.1\linewidth]{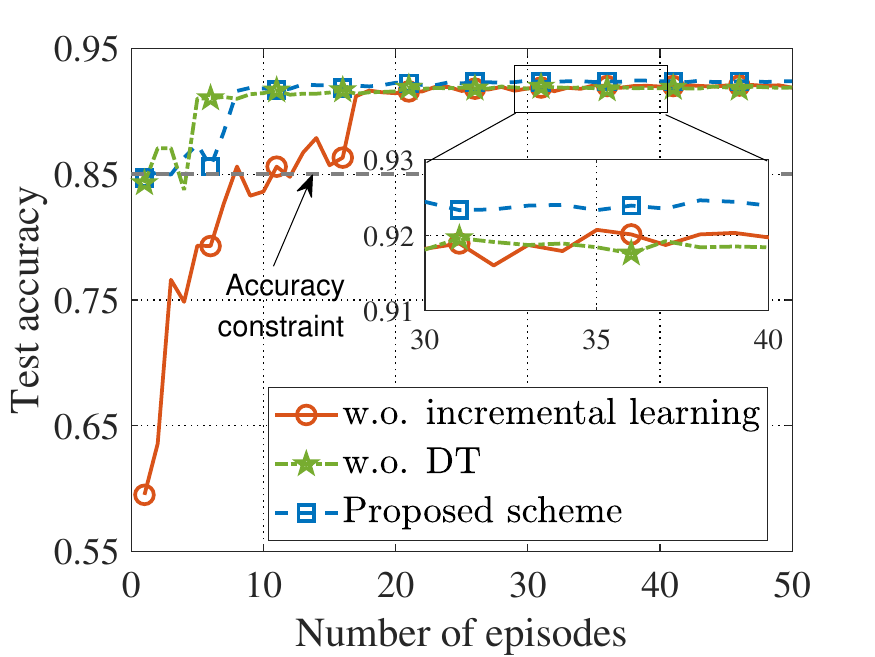}
		\caption{Model training iterative curve}
		\label{Sim1}
	\end{subfigure}
	\centering
	\begin{subfigure}{0.325\linewidth}
		\centering
		\includegraphics[width=1.1\linewidth]{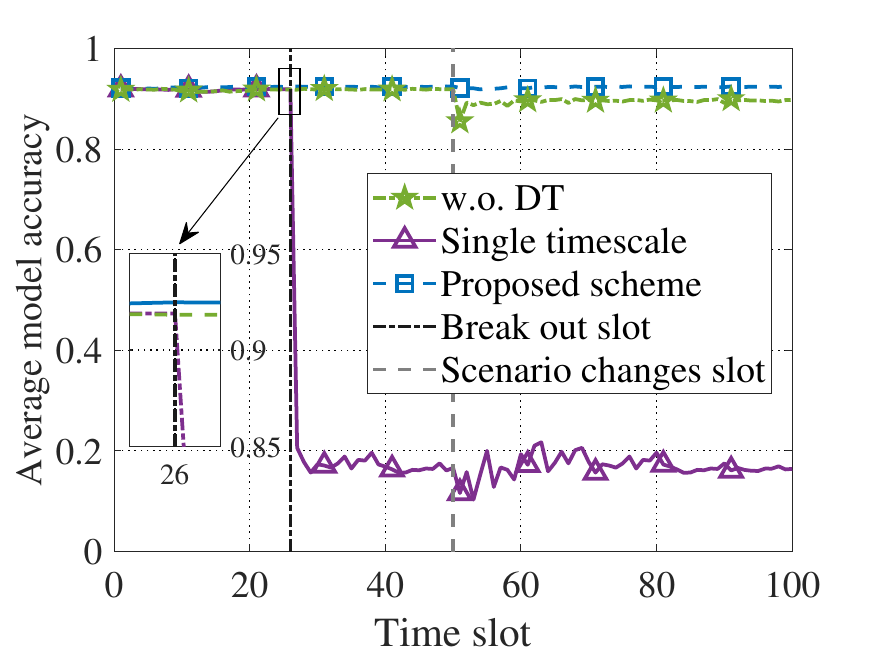}
		\caption{Model accuracy}
		\label{Sim2}
	\end{subfigure}
        \hspace{0.1mm}
	\centering
	\begin{subfigure}{0.325\linewidth}
		\centering
		\includegraphics[width=1.1\linewidth]{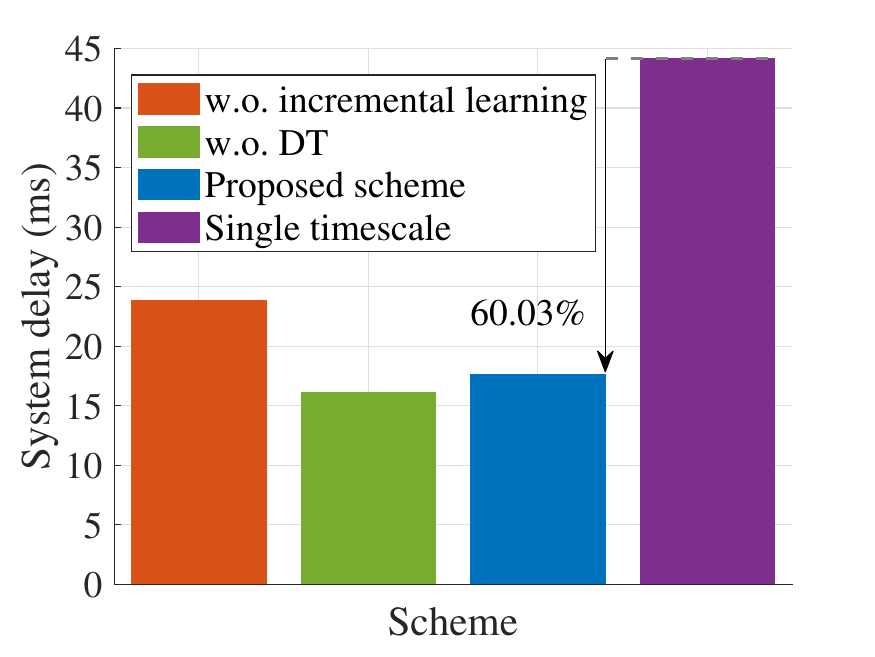}
		\caption{System delay}
		\label{Sim3}
	\end{subfigure}
	\caption{Performance of the proposed scheme compared with baselines on CIFAR-10 datasets.}
	\label{da_sim}
\end{figure*}

In Fig.~\ref{da_sim}(a), the iterative model training curve per episode is depicted, showing that the proposed DT-based incremental learning approach outperforms both the approaches without DT and without incremental learning scheme. Additionally, the use of DT-generated data incurs additional training costs but results in higher model accuracy. All schemes can be retrained into a DNN model that meets accuracy constraints effectively. In Fig.~\ref{da_sim}(b), when a specific probability of images is present in the test set, the single timescale results in abrupt breakouts, while the proposed two-timescale scheme can effectively prevent such occurrences. It can be observed that with our proposed approach, calculating average accuracy and making retraining decisions in each large timescale increases system robustness. For schemes without DT, due to absent training sets for incremental learning methods compared to well-trained models at initialization, accuracy decreases even with updates model in another large timescale, especially when data distribution varies greatly. In Fig.~\ref{da_sim}(c), we observe the average system delay over 100 time slots. It is clear that our proposed DT-assisted approach reduces total system delay by more than 60\% compared to single timescale schemes. Conversely, without incremental learning to adapt to data changes and completely retrain models, system delay are not appropriately adjusted.

\section{Conclusion}
In this paper, we have proposed a two-timescale scheme for network resource allocation assisted by DT and incremental learning to minimize the system delay. The system architecture is divided into physical and DT systems, with the physical scenarios providing real-time user information and the DT system providing additional data and making decisions on model retraining. Experimental results demonstrate that, compared to other schemes, the proposed scheme can significantly reduce system delay while increasing average model accuracy, which can be exploited to image identification and autonomous driving. For the future work, challenges arise in balancing computation resource cost and retraining accuracy under system delay constraints due to data generation and statistical information learning in DT, requiring further investigation.


\bibliographystyle{IEEEtran}

\bibliography{CJYref.bib}

\end{document}